# High-$T_c$ superconductivity in FeSe at high pressure: Dominant hole carriers and enhanced spin fluctuations

J. P. Sun[1], G. Z. Ye[1,2], P. Shahi[1], J.-Q. Yan[3], K. Matsuura[4], H. Kontani[5], G. M. Zhang[6], Q. Zhou[2], B. C. Sales[3], T. Shibauchi[4], Y. Uwatoko[7], D. J. Singh[8#], and J.-G. Cheng[1#]

[1]Beijing National Laboratory for Condensed Matter Physics and Institute of Physics, Chinese Academy of Sciences, Beijing 100190, China
[2]School of Physical Science and Astronomy, Yunnan University, Kunming 650091, China
[3]Materials Science and Technology Division, Oak Ridge National Laboratory, Oak Ridge, TN 37831, USA
[4]Department of Advanced Materials Science, University of Tokyo, Kashiwa, Chiba 277-8561, Japan
[5]Department of Physics, Nagoya University, Furo-cho, Nagoya 464-8602, Japan
[6]State Key Laboratory of Low Dimensional Quantum Physics and Department of Physics, Tsinghua University, Beijing 100084, China
[7]The Institute for Solid State Physics, University of Tokyo, Kashiwa, Chiba 277-8581, Japan
[8]Department of Physics and Astronomy, University of Missouri, Columbia, Missouri 65211-7010, USA
[#]E-mails: jgcheng@iphy.ac.cn, singhdj@missouri.edu

## Abstract

The importance of electron-hole interband interactions is widely acknowledged for iron-pnictide superconductors with high transition temperatures ($T_c$). However, high-$T_c$ superconductivity without hole carriers has been suggested in FeSe single-layer films and intercalated iron-selenides, raising a fundamental question whether iron pnictides and chalcogenides have different pairing mechanisms. Here, we study the properties of electronic structure in the high-$T_c$ phase induced by pressure in bulk FeSe from magneto-transport measurements and first-principles calculations. With increasing pressure, the low-$T_c$ superconducting phase transforms into high-$T_c$ phase, where we find the normal-state Hall resistivity changes sign from negative to positive, demonstrating dominant hole carriers in striking contrast to other FeSe-derived high-$T_c$ systems. Moreover, the Hall coefficient is remarkably enlarged and the magnetoresistance exhibits anomalous scaling behaviors, evidencing strongly enhanced interband spin fluctuations in the high-$T_c$ phase. These results in FeSe highlight similarities with high-$T_c$ phases of iron pnictides, constituting a step toward a unified understanding of iron-based superconductivity.

# Introduction

Obtaining a unified understanding of unconventional superconductivity (SC) is one of the key research goals in modern condensed matter physics. The discovery of iron-based superconductors (FeSCs) greatly enlarged the space of available materials for investigation of unconventional SC providing new avenues for approaching the long-standing mystery of unconventional SC.[1] SC in the FeSCs generally emerges on the border of antiferromagnetically ordered states, similar to cuprates and heavy Fermions, which suggests possible magnetic mechanisms.[2] Unlike cuprates, however, the parent compounds of FeSCs are generally antiferromagnetic metals rather than Mott insulators. The Fermi surface (FS) topology and its interplay with magnetism have been considered key ingredients in understanding the superconducting mechanism of the FeSCs.[3,4]

The FS of the FeAs-based superconductors typically consists of hole-like pockets near the Brillouin zone center (Γ point) and electron-like pockets near the Brillouin zone corners (M point). This has led to the proposal that inter-band scattering between the hole and electron pockets provides the mechanism for electron pairing in the FeSCs, leading to an s± pairing state favored by the antiferromagnetic fluctuations.[3,4] This picture, however, is challenged by the distinct FS topology observed in the FeSe-derived high-$T_c$ (> 30 K) superconductors, including $A_xFe_{2-y}Se_2$ (A = K, Cs, Rb, Tl),[5] $Li_x(NH_2)_y(NH_3)_{1-y}Fe_2Se_2$, (Li,Fe)OHFeSe,[6] and monolayer FeSe film on $SrTiO_3$,[7] in which only the electron pockets are observed near the Fermi level from the high-resolution ARPES experiments. A recent study on the gate-voltage-regulated FeSe flakes further supports the conclusion that electron doping plays a crucial role in achieving a high $T_c$ in the FeSe-based materials.[8] Thus, the distinct FS topology between the FeAs- and FeSe-based materials has challenged current theories on a unified understanding on the mechanism of FeSCs.

Bulk FeSe is a compensated semimetal with both electron and hole FS similar to the FeAs based materials, but without antiferromagnetism and with a low $T_c$. In contrast to electron-doping approaches, the application of high pressure, does not chemically introduce extra electron carriers in the system, yet can still enhance $T_c$ of bulk FeSe up to ~40 K near 6 GPa. More importantly, our recent high-pressure resistivity study has shown explicitly that the optimal $T_c$ is achieved when the long-range magnetic order just vanishes,[9] Fig. 1, reminiscent of the situations seen frequently in the FeAs-based superconductors. However, to make this connection, it is important to have information about the evolution of the FS under high pressure – a regime in which ARPES experiments are impractical, and where quantum oscillation measurements are challenging.

Here we report Hall resistivity $\rho_{xy}$ measurements under hydrostatic pressures up to 8.8 GPa in order to gain insights into the electronic structure evolution of FeSe at high pressure. Our results demonstrate that the electrical transport properties of FeSe at high pressures with $T_c^{max}$ = 38.3 K are dominated by the hole carriers, Fig. 1, which is in striking contrast with the known FeSe-derived high-$T_c$ superconductors that are usually heavily electron doped. The results are in accord with density functional calculations. In addition, we observed an enhancement of Hall coefficient $R_H$ near the critical pressure where the optimal $T_c$ is realized with a simultaneous suppression of the long-range magnetic order. This implies a strong reconstruction of the Fermi surface due to antiferromagnetic order, consistent with the ordering pattern driven by interband scattering, and consistent with density functional calculations. Importantly, our results show a continuous path to high temperature superconductivity in chalcogenides without electron doping, making a strong connection between the arsenides and chalcogenides.

## Experimental details

High-quality FeSe single crystals used in the present study were grown by the flux method. Detailed characterization of these crystals have been carried out in our previous studies on the high-pressure resistivity measurements.[9] Hall resistivity $\rho_{xy}(H)$ measurements at various temperatures were performed under hydrostatic pressures up to 8.8 GPa with the "palm" cubic anvil cell (CAC) in the Institute of Physics, Chinese Academy of Sciences. The current was applied within the *ab* plane and the magnetic field applied along the *c* axis. The $\rho_{xy}(H)$ ($\rho_{xx}(H)$) data were anti-symmetrized (symmetrized) with respect to the data collected between +5 and -5 T. The pressure values inside the CAC were calibrated at room temperature by observing the characteristic transitions of bismuth. Glycerol was employed as the pressure transmitting medium.

## Density Functional Calculations

Our density functional calculations were performed using the general potential linearized augmented planewave (LAPW) method as implemented in the WIEN2k code. Our calculations are for magnetically ordered stoichiometric FeSe. For the structure we performed polynomial fits of existing diffraction data for lattice parameter variations under pressure in both the tetragonal and orthorhombic phases.[10] The electronic structure depends on the height of the Se above the Fe plane, but sufficiently accurate high pressure experimental data for this parameter is unavailable. As such, we determined the Se height by relaxing the Se position, with stripe antiferromagnetic order in the orthorhombic structure. For this purpose we used the PBE GGA functional. The resulting Se height was then held fixed for all the calculations at that pressure, consistent with the known behavior of the FeSCs[11] using the BoltzTraP code[12], with dense Brillouin zone sampling and the constant scattering time approximation.[13] We note that the constant scattering time approximation is valid away from phase transitions that produce additional scattering, such as the nematic and antiferromagnetic transitions of FeSe. The calculations presented are for 200 K, which is well above these transitions.

## Results and discussions

The electronic structure and FS topology of bulk FeSe and FeSe-derived high-$T_c$ materials have been extensively studied by ARPES [14, 15-17] and/or quantum oscillations[17-20]. For the bulk FeSe, its FS in the tetragonal phase is similar to those of typical iron-pnictides superconductors, consisting of hole-like FS around the Γ point and electron-like FS around the M point, in general accord with density functional calculations [21,22], although with a significant mass enhancement. A significant FS reconstruction takes place near the structural transition at $T_s \approx 90$ K, manifested by a dramatic splitting of $d_{yz}/d_{xz}$ bands around both Γ and M points.[14,15] The splitting of ~50 meV is too large to be explained solely by a structural transition, but has to be attributed to the development of an electronic nematicity. As a result, the FS in the orthorhombic phase consists of one hole and two electron pockets with very tiny carrier numbers, *i.e.* < 0.01 carriers/Fe.[18] The comparable size of Fermi energies $E_F$ and the superconducting energy gap $\Delta$ in such a low-carrier system may make FeSe a unique superconductor approaching the BCS-BEC crossover regime.[23] In contrast, APRES measurements on the monolayer and interacted FeSe-derived high-$T_c$ (> 30K) superconductors revealed a common FS topology featured by the electron pockets only near the M point.[24]

In addition to the ARPES and quantum oscillations, Hall data also carries useful information about the electronic structure. The Hall resistivity $\rho_{xy}$ at ambient pressure displays characteristic evolution upon cooling.[19,25] For $T > T_s$, $\rho_{xy}(H)$ curves are all linear, but its slope or the Hall coefficient $R_H$ changes sign twice with decreasing temperature due to a slight variation of mobility for the almost compensated electron and hole carriers, as shown in Fig. 3a. A moderate enhancement of $R_H$ is evidenced near $T_s$, below which a strong nonlinearity of $\rho_{xy}(H)$ develops with an initial negative slope followed by a positive one at higher field.

With this information in hand, we proceed to the evolution of the Hall resistivity under high pressure. The field dependence of Hall resistivity $\rho_{xy}(H)$ at various temperatures under different pressures are displayed in Fig. 2. As seen in Fig. 2a and 2b, the $\rho_{xy}(H)$ data at 1.5 and 1.8 GPa share many similar features as those at ambient pressure. Specifically, $\rho_{xy}(H)$ curves are linear at $T > 40$ K and the slope changes sign twice from positive to negative and then back to positive upon cooling from room temperature, in accordance with the compensated semimetal character. A nonlinearity develops for $\rho_{xy}(H)$ curves below 40 K and the initial slope eventually becomes negative for $T < 30$ K, but tends to change sign again under higher magnetic field. According to the previous studies at ambient pressure, the low-field negative slope has been ascribed to the emergence of the minority electron carriers with high mobility.[19] When the pressure is increased to 3.8 GPa, surprisingly, all $\rho_{xy}(H)$ curves exhibit a positive slope without any temperature-induced sign reversal in the whole temperature range, Fig. 2c, implying that the hole carriers become dominate. The positive slope increases gradually with decreasing temperature to 40 K, below which a nonlinearity also appears, but the initial slope remains positive down to the superconducting transition temperature $T_c$. Such a hole dominated Hall effect was observed to persistent up to 8.8 GPa, the highest pressure in this study. As seen in Fig. 2d-2f, two features are noteworthy in this pressure range. At the first place, the linear $\rho_{xy}(H)$ at high temperatures is replaced gradually by a slightly nonlinear, concave behavior upon cooling. At 6.3 GPa, such a nonlinearity of $\rho_{xy}(H)$ is found to persist up to 100 K. In the high-$T_c$ cuprates, the develop of such nonlinearity of $\rho_{xy}(H)$ above $T_N$ has been attributed to the two-dimensional antiferromagnetic spin fluctuations. Secondly, with increasing pressure the $\rho_{xy}(H)$ at a given temperature is found to be first enhanced profoundly and then decrease quickly. For example, $\rho_{xy}$ at 5 T and 40 K first increases from ~30× $10^{-9}$ Ω m at 3.8 GPa to ~55 × $10^{-9}$ Ω m at 4.8 and 6.3 GPa and then decreases to 25 × $10^{-9}$ Ω m at 7.8 GPa, Fig. 2c-2f. As discussed below, the enhancement of Hall resistivity correlates intimately with the AF fluctuations. Regardless of these details, the immediate message that we can learn from the above Hall resistivity is that the electronic structure of FeSe undergoes a dramatic change under pressure, making the hole carriers dominating the electronic transport under high pressures above 3 GPa.

Our DFT calculations show five sheets of Fermi surface in the non-magnetic tetragonal structure. The calculated Fermi surface at 6 GPa is shown in Fig. 4a. There are three hole cylinders, containing 0.011, 0.160 and 0.191 holes per two FeSe formula unit cell, respectively, and two compensating electron cylinders with 0.209 and 0.153 electrons, respectively. The smallest hole cylinder closes off and becomes a three dimensional ellipsoid between 8 GPa and 10 GPa, leading to a change in sign of the Hall number for field along the $z$-direction. We note that the evolution of the Fermi surface with pressure up to 10 GPa is smooth, and that there are no changes in topology or additional surfaces, except for this closing off of the smallest hole cylinder. The DFT calculations also show an instability against the SDW type antiferromagnetic order, accompanied by reconstruction of the FS. This reconstruction removes most of the FS. This is the case both for GGA and for local spin density approximation (LSDA) calculations and in both cases the magnetic tendency is overestimated compared with experiment, qualitatively similar to the case of the FeAs-based superconductors, and presumably reflecting a large renormalization by spin fluctuations[11]. At 6 GPa the LSDA AF state has a density of states that is 0.32 of the non-spin-polarized case with a calculated moment of 1.1 $\mu_B$/Fe.

In order to gain detailed insight on the evolution of the Hall effect as functions of temperature and pressure, we have plotted in Fig. 3 the temperature dependence of Hall coefficient, defined as the field derivative of $\rho_{xy}$, $R_H \equiv d\rho_{xy}/dH$, at the zero-field limit, together with the zero-field resistivity curve $\rho(T)$ at each pressure. Field is applied along the $c$-axis. As shown in Fig. 3a, $R_H$ at ambient pressure is small for $T > 100$ K, within the range of ± 0.5 × $10^{-9}$ m$^3$/C, and changes sign twice upon cooling. A moderate enhancement of $R_H$ to ~ 2 × $10^{-9}$ m$^3$/C is evidenced towards $T_s$, below which $R_H$ reverses sign again and exhibits a strong tendency to large negative values. Such a dramatic change of Hall coefficient at $T_s$ corresponds to the FS reconstruction due to the formation of electronic nematicity as mentioned above. As shown in Fig. 3b, $R_H(T)$ at 1.5 GPa displays very similar behaviors as that at ambient pressure, except that $T_s$ has been shifted down to ~ 50 K. When the pressure is increased to 1.8 GPa, the nematic order almost vanishes and the long-range antiferromagnetic order starts to emerge at $T_m \approx 20$ K, which is manifested as an upturn anomaly in resistivity curve, Fig. 3c. Although the overall behaviors of $R_H(T)$ at 1.8 GPa resemble that at 1.5 GPa, including the magnitude and the twice sign reversals above 40 K, the negative Hall coefficient $R_H(T)$ at low temperatures exhibits a much steeper growth upon cooling below $T_m$ with respect to those observed below $T_s$ at 1.5 GPa. This means that a more pronounced FS reconstruction takes place at the AF order, which removes a larger portion of FS. Similar results have been reported recently by Terashima *et al.* through measurements of Shubnikov-de Haas quantum oscillations[26] and Hall effect under pressures[27]. Despite of this difference, the major characteristics of FS topology are not expected to change dramatically up to 2 GPa given the similarity of Hall effects; the normal state above $T_m$ is characterized by a compensated semimetal while that below $T_m$ is likely dominated by the minority electron carriers with high mobility.

The situation changes dramatically when the pressure is increased to above 3 GPa. As seen in Fig. 3d for $P = 3.8$ GPa, the AF order at $T_m \approx 40$ K is manifested as a pronounced drop in resistivity, and the positive $R_H(T)$ experiences a noticeable enhancement when approaching $T_m$ from above. It is interesting to note that the enhancement of $R_H$ starts around a characteristic temperature $T^* \sim 150$ K near which the resistivity curve exhibits a clear upward deviation from the quasi linear-in-T behavior at high temperatures. Above $T^*$, $R_H$ is nearly temperature independent and takes tiny values around $0.5 \times 10^{-9}$ m$^3$/C. Then, $R_H(T)$ decreases quickly below $T_m$ until reaching zero around $T_c$. Nearly identical features are observed at 4.8 GPa, except that the enhancement at $T_m$ becomes stronger as seen in Fig. 3e; $R_H$ at $T_m$ increases considerably from $\sim 6 \times 10^{-9}$ m$^3$/C at 3.8 GPa to $\sim 14 \times 10^{-9}$ m$^3$/C at 4.8 GPa. When the pressure is further increased to 6.3 GPa, the AF order becomes destabilized and the superconducting transition temperature reaches the highest $T_c^{max} = 38.3$ K. As seen in Fig. 3f, nearly T-independent $R_H$ above $T^* \sim 150$K is greatly enhanced upon cooling until approaching $T_c$; the enhancement is much more pronounced with the maximum $R_H$ reaching $\sim 20 \times 10^{-9}$ m$^3$/C. Similarly, an upward deviation from the linear-T behavior of resistivity was also observed around $T^*$. According to our previous magneto-resistivity measurements, the magnetic order still exists at 6.3 GPa, but takes place at $T_m$ slightly lower than $T_c$.

Upon further increase of pressure to 7.8 and 8.8 GPa, the magnetic order vanishes and $T_c$ decreases slightly. As seen in Fig. 3g and 3h, the normal-state resistivity displays a nearly perfect linear-in-T dependence in a wide temperature range from room temperature down to $T_c$. Concomitantly, the enhancement of $R_H(T)$ becomes much weakened and tends to diminished at 8.8 GPa, suggesting that the carrier density or the density of states at Fermi level is greatly enhanced. These observations suggest that the enhancement of $R_H$ correlates intimately with the deviation of resistivity from the high-temperature linear-in-T behaviors.

Our constant scattering time approximation calculations based on the DFT electronic structure show a positive Hall coefficient (Fig. 4b) for $H // c$, consistent with experimental results, and in addition show an increase up to 6 GPa and then a decrease above 8 GPa. This closely follows the experimental observation at high temperatures, making a connection between the smooth evolution of the electronic structure with pressure in the DFT calculations and the experimental evolution of the Hall data. This supports the conclusion that the electronic structure of FeSe remains similar to the FeAs superconductors with compensating zone-center hole sheets and zone-corner electron sheets including pressures where $T_c$ is high. It is also noteworthy that the Hall coefficient is predicted to have opposite sign for field in the *ab* plane, as is possible in compensated metals, but not usually in semiconductors. This underscores the metallic nature of the compound, different from a doped insulator.

The major findings of the present study can be summarized in the phase diagram superimposed with the contour plot of Hall coefficient $R_H$. As seen in Fig. 1, the electronic structure of FeSe in the normal state just above $T_c$ undergoes a dramatic reconstruction under pressure; it changes from the dominated electron-type to hole-like around 2 GPa where the nematic order is just suppressed with an concomitant emergence of AF order. The observation of dominant hole carriers at $P > 3$ GPa, especially in the pressure range where high-$T_c$ superconductivity can be achieved is surprising in that the hole FSs are missing in all the known FeSe-derived high-$T_c$ superconductors. In addition, the stabilization of the AF order under high pressure can be attributed to the presence of hole pockets that admits FS nesting mechanisms for selecting the AF order. Although the current study puts FeSe at variance with other FeSe-derived high $T_c$ superconductors, the observation of high-$T_c$ superconductivity on the border of AF order, in reminiscent of the FeAs-based superconductors, suggests that the electron-hole interband interactions are important for both FeSe under pressure and the FeAs-based superconductors.

The presence of significant AF spin fluctuations is supported by the dramatic enhancement of $R_H$ above $T_m$ centered near 6.5 GPa, as seen in Fig. 1. Since the carrier number is not expected to change dramatically in the paramagnetic states, the enhancement of $R_H$ upon cooling has to be associated with the increased scattering rate due to AF spin fluctuations. Indeed, the Curie-Weiss-like temperature dependence of $R_H$ has been observed in cuprate, heavy-fermion, and iron-pnictide superconductors with strong AF spin fluctuations.[28,29] In the theories involving the vertex corrections[30,31], where the backflow effect originating from the charge conservation law in the presence of strong electron correlations is taking into account[28], the enhanced $R_H$ in the presence of AF spin fluctuations can be naturally understood by the enlarged AF correlation length. It has been also pointed out theoretically and observed experimentally that such strong AF spin fluctuations also affect the magnetoresistance $\Delta\rho_{xx}(H)/\rho_{xx}(0) = [\rho_{xx}(H)-\rho_{xx}(0)]/\rho_{xx}(0)$; unlike the conventional Kohler's rule (the scaling by $H/\rho_{xx}(0)$), the magnetoresistance can be scaled by $\tan^2\Theta_H$ where $\Theta_H$ is the Hall angle. As demonstrated in Fig. 5, this violation of Kohler's rule and the modified scaling with the Hall angle are observed for our magnetoresistance data at 6.3 GPa, where the enhancement of $R_H$ is the strongest. When the magnetic order takes place below $T_m$, the scattering is reduced significantly, leading to the observed resistivity drop in Fig. 3d-3e. It should be noted that recent X-ray and NMR measurements up to ~3GPa suggest the first-order nature of field-induced magnetic transition.[32,33] The present results nevertheless show that the AF spin fluctuations are strongly enhanced near the magnetic phase boundary, where the high-$T_c$ superconducting phase appears. An alternative view of the Hall coefficient in the 3D plot of $R_H(P, T)$ in Fig. S1 indicated that the $R_H(T)$ at 6.3 GPa tends to diverge if the superconducting transition is absent, which implies that they might exist a hidden magnetic quantum critical point near $P_c$. The observed linear-in-T resistivity in a large temperature range for $P > P_c$, Fig. 3(g, h), is also consistent with such a scenario.

Therefore, our current high-pressure magneto-transport study underscores the importance of interband spin fluctuations in achieving high-$T_c$ superconductivity in FeSe under pressure. The stabilization of AF order above 2 GPa is likely associated with the FS reconstruction. This is consistent with a FS driven magnetic order similar to scenarios discussed for the FeAs materials. Indeed, very recent theoretical calculations[34] show that the increase of the relative Se height induces the $d_{xy}$ hole FS pocket (one of the 3 hole pockets in Fig. 4a), which results in the improvements of the intra-orbital interband nesting and thus promotes the stripe-type AF order. When the AF order is destabilized by the application of high pressure, the AF spin fluctuations may play an essential role for achieving high-$T_c$ superconductivity as found in the FeAs-based superconductors.

Finally, we comment on the high-$T_c$ phases in single-layer films and intercalates of FeSe. The hole pockets at the Γ point may vanish as the Fermi level is raised by electron doping in these materials as reported by ARPES experiments. However, recent experiments in $K_xFe_{2-y}Se_2$ report the presence of the hole band near the Fermi level that has been hidden possibly due to matrix elements effects in ARPES measurements[34]. Moreover, even when the hole band is slightly away from the Fermi level, the interband interactions may still be important in the superconductivity[35]. Recently, Linscheild *et al.*[36] proposed that the incipient hole band that is 50-100 meV below the Fermi level may contribute significantly to the spin-fluctuation pairing in the strong coupling regime and can lead to a high $T_c$. Thus, our present work, together with these recent proposals, can be considered as a significant step forward in making a unified picture on the current understanding of FeSCs, specifically by demonstrating that high $T_c$ in FeSe can be achieved with an electronic structure and other characteristics similar to the FeAs-based high-$T_c$ superconductors.

## Acknowledgements
We thank Dr. T. Xiang, Dr. X. J. Zhou, Dr. J. L. Luo, and Dr. Kui Jin for very helpful discussions. This work is supported by the National Basic Research Program of China (Grant No.2014CB921500), National Science Foundation of China (Grant No. 11574377), the Strategic Priority Research Program and Key Research Program of Frontier Sciences of the Chinese Academy of Sciences (Grant Nos. XDB07020100, QYZDB-SSW-SLH013), and the Opening Project of Wuhan National High Magnetic Field Center (Grant No. 2015KF22) Huazhong University of Science and Technology. J.-Q. Y. and B.C.S. are supported by the US Department of Energy, Office of Science, Basic Energy Sciences, Materials Sciences and Engineering Division. Work in Japan is supported by Grants-in-Aid for Scientific Research (KAKENHI) from Japan Society for the Promotion of Science.

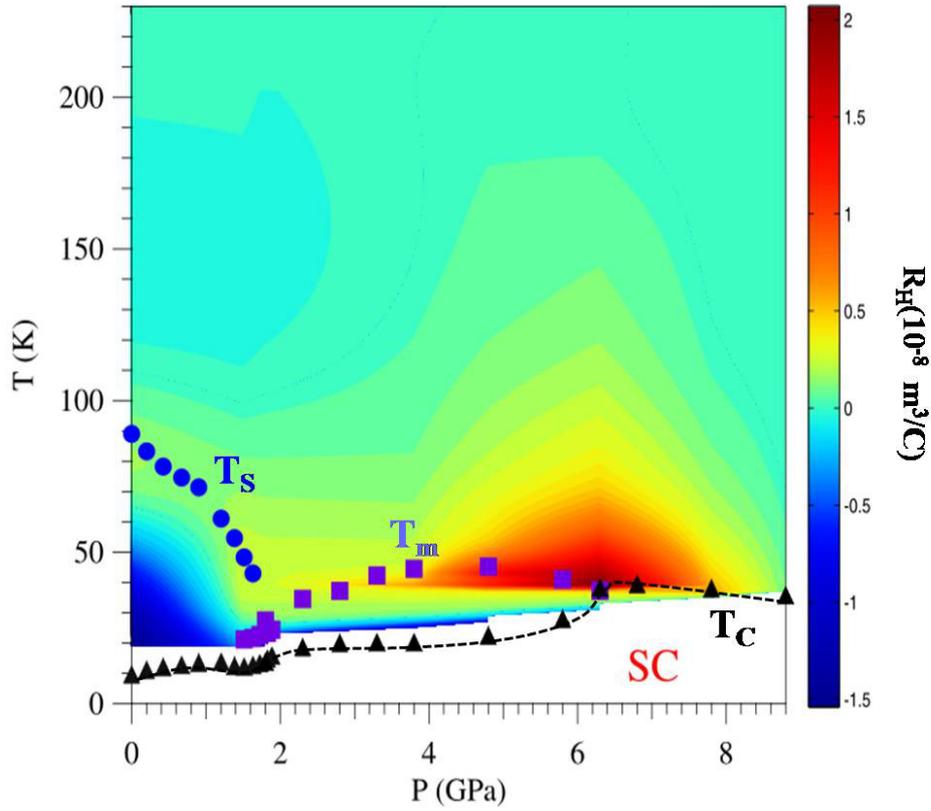

FIG. 1: **Phase diagram and Hall coefficient of FeSe.** Temperature-pressure phase diagram of FeSe is superimposed by a contour plot of Hall coefficient $R_H$. The structural (nematic) transition ($T_s$), pressure-induced magnetic transition ($T_m$) and superconducting (SC) transition ($T_c$) have been determined by the resistivity measurements under pressure[9]. The dashed curve is a guide for eyes.

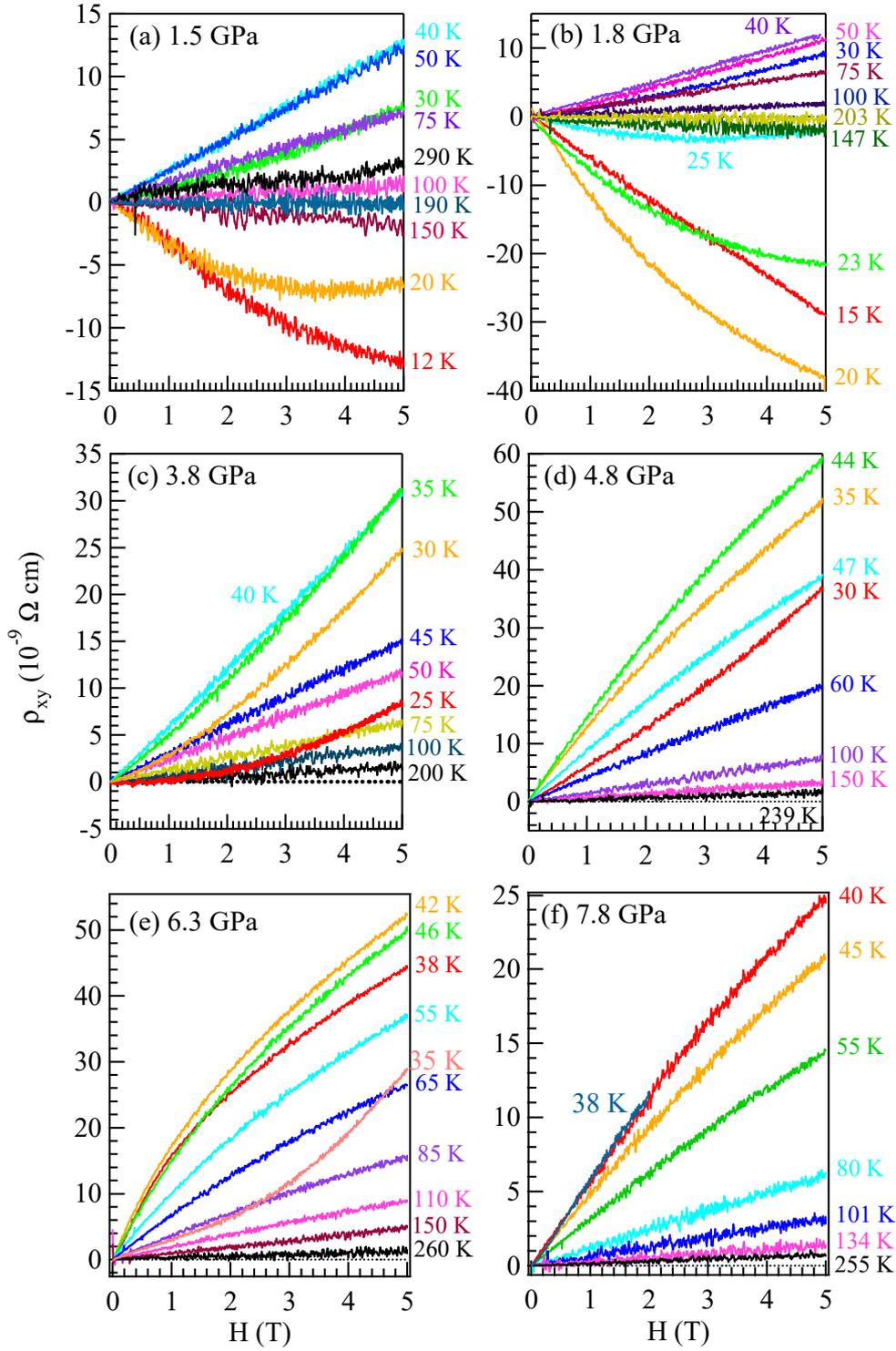

FIG. 2: **Normal-state Hall resistivity of FeSe.** Field dependence of Hall resistivity $\rho_{xy}(H)$ of FeSe single crystal measured under various temperatures at different pressures.

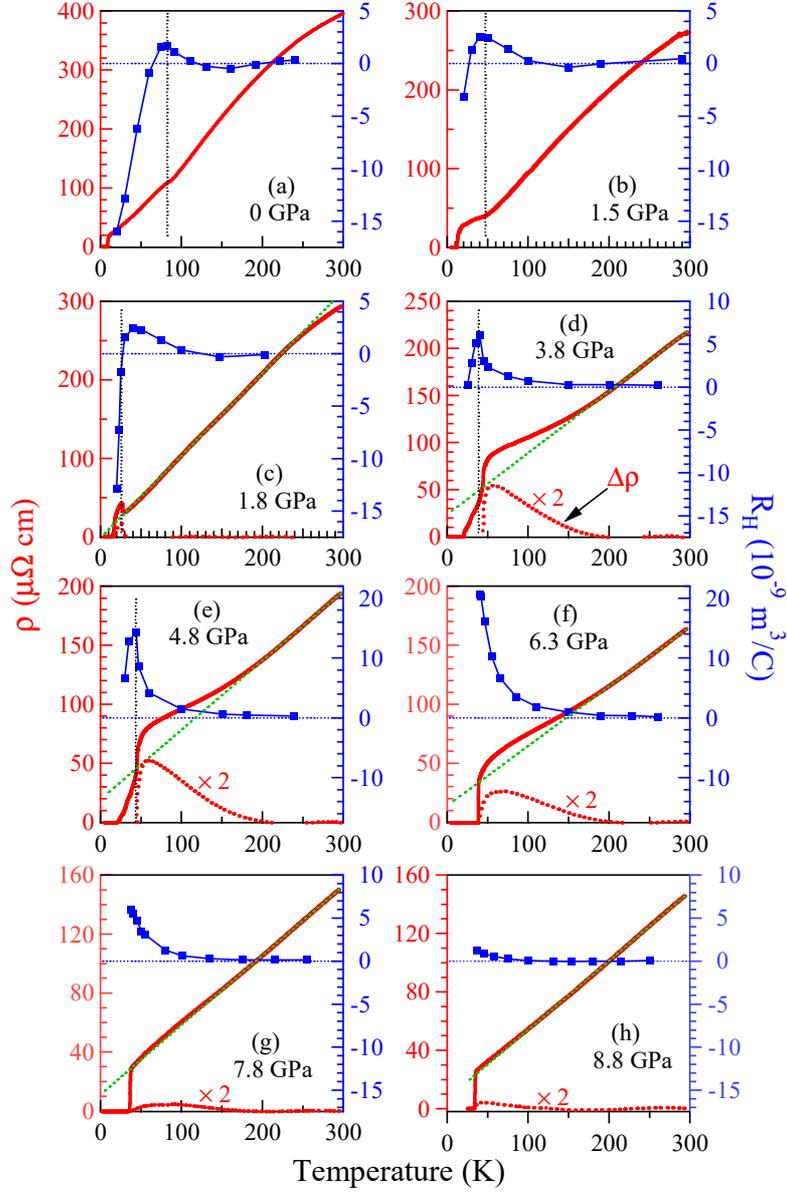

FIG. 3: **Resistivity and Hall coefficient vs. temperature of FeSe.** Temperature dependence of Hall coefficient $R_H$ (blue, right axis) defined as the field derivative of $\rho_{xy}$, $R_H \equiv d\rho_{xy}/dH$, at the zero-field limit, together with the zero-field resistivity curve $\rho(T)$ (red, left axis) at each pressure. The vertical dotted lines in (a-e) mark the nematic order transition at $T_s$ and the magnetic order at $T_m$; the horizontal dotted lines in all figures indicated the zero $R_H$. The resistivity difference $\Delta\rho$ (scaled by a factor of 2) was obtained by subtracting from the measured resistivity $\rho(T)$ the linear-fitting curve at high temperatures as indicated by the broken line (green).

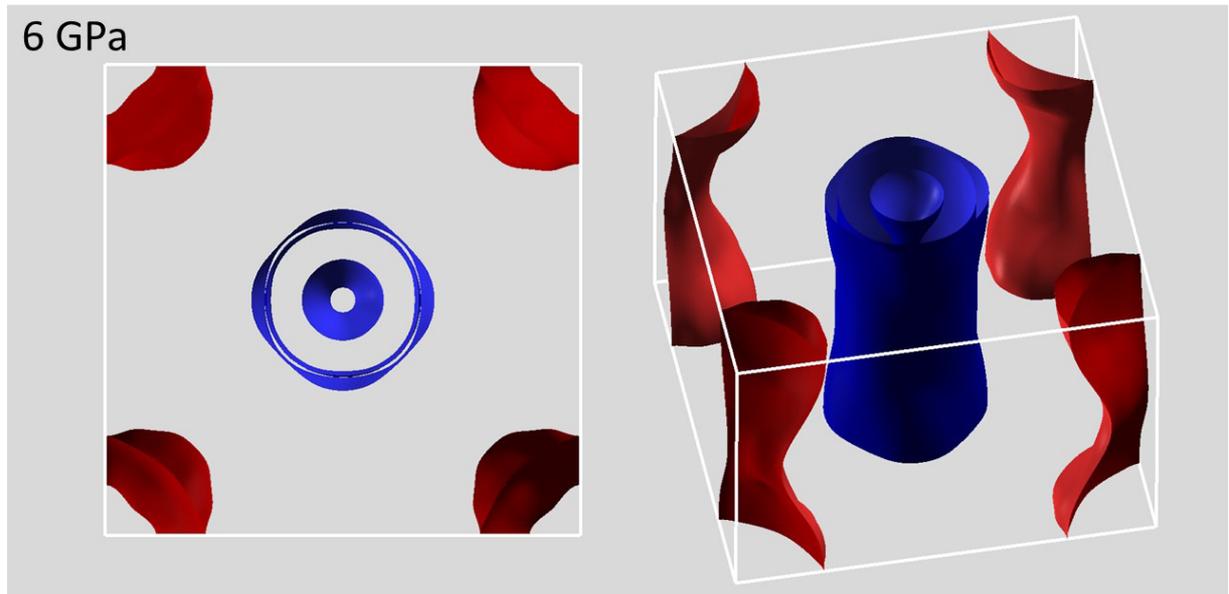

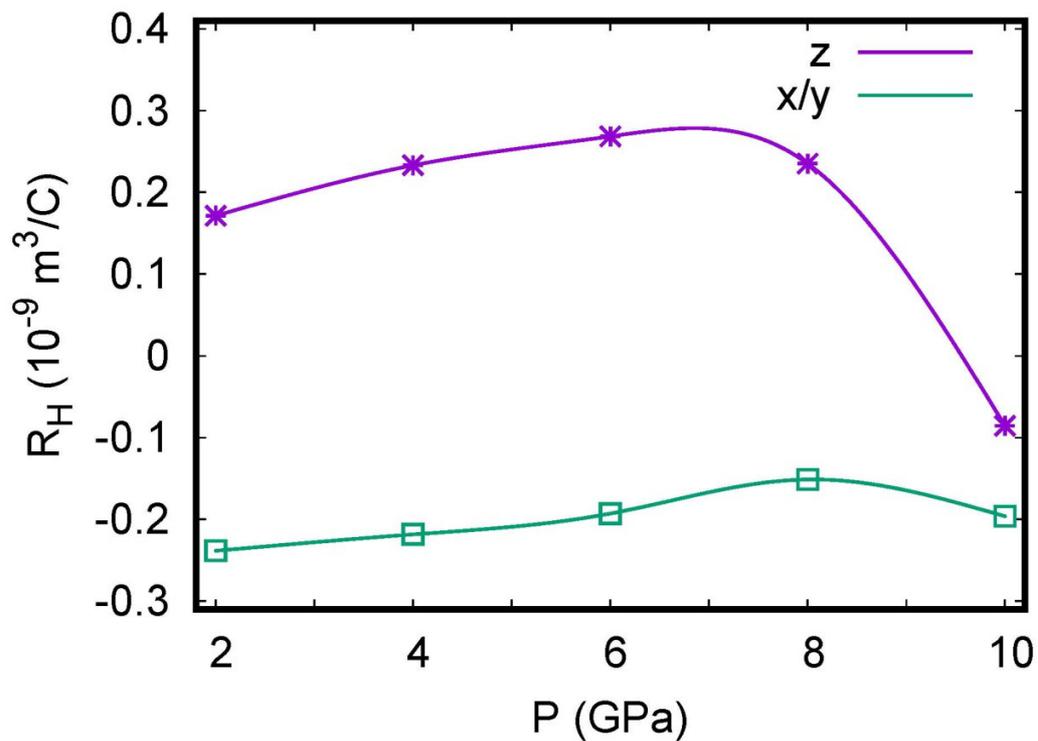

Fig. 4: **Calculated Fermi surface and high-temperature Hall coefficient for FeSe. (a)** Results of first-principles calculations for Fermi surface of tetragonal FeSe at 6 GPa, viewed along the *c*-axis (left) and at an angle (right), with hole bands shown in blue and electron bands shown in red. **(b)** Calculated 200 K Hall coefficient as a function of pressure for field along the *c* axis (z, $R_{xy}$) corresponding to the experimental geometry, and for field in plane (x/y, $R_{xz}$).

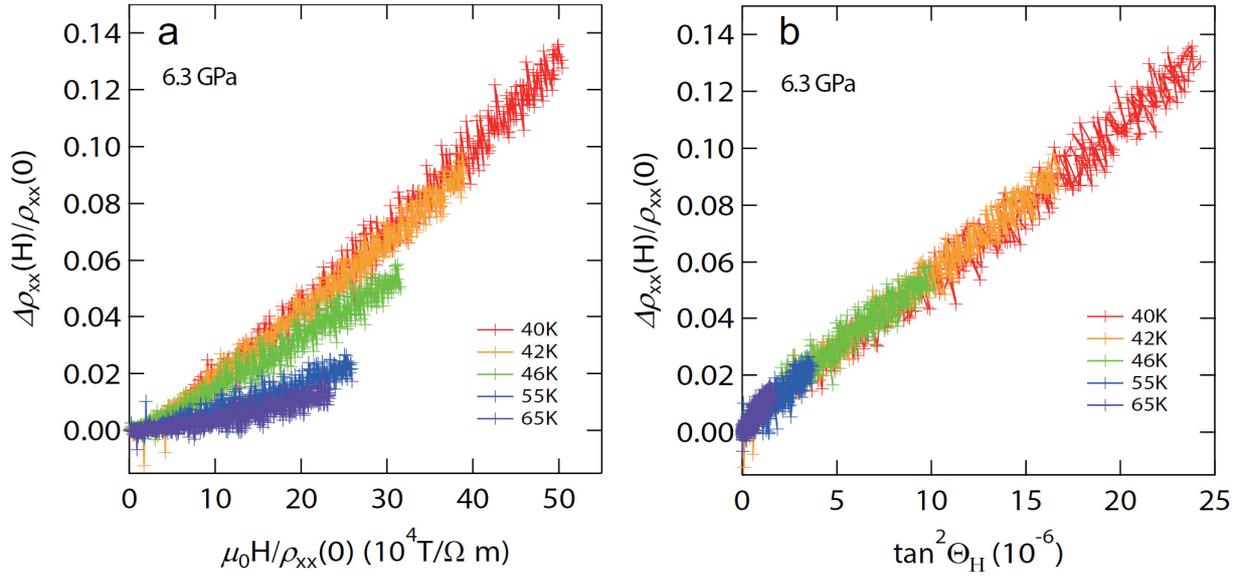

Fig. 5: **Normal-state magnetoresistance at 6.3 GPa. (a)** Magnetoresistance at different temperatures are plotted against magnetic field divided by zero-field resistivity at each temperature (Kohler's plot). The deviations indicate the violation of the Kohler's rule. **(b)** The same data plotted against $\tan^2\Theta_H$.

# Supporting materials

# High-$T_c$ superconductivity in FeSe at high pressure: Dominant hole carriers and enhanced spin fluctuations

J. P. Sun[1], G. Z. Ye[1,2], P. Shahi[1], J.-Q. Yan[3], K. Matsuura[4], H. Kontani[5], G. M. Zhang[6], Q. Zhou[2], B. C. Sales[3], T. Shibauchi[4], Y. Uwatoko[7], D. J. Singh[8#], and J.-G. Cheng[1#]

[1]Beijing National Laboratory for Condensed Matter Physics and Institute of Physics, Chinese Academy of Sciences, Beijing 100190, China
[2]School of Physical Science and Astronomy, Yunnan University, Kunming650091, China
[3]Materials Science and Technology Division, Oak Ridge National Laboratory, Oak Ridge, TN 37831, USA
[4]Department of Advanced Materials Science, University of Tokyo, Kashiwa, Chiba 277-8561, Japan
[5]Department of Physics, Nagoya University, Furo-cho, Nagoya 464-8602, Japan
[6]State Key Laboratory of Low Dimensional Quantum Physics and Department of Physics, Tsinghua University, Beijing 100084, China
[7]The Institute for Solid State Physics, University of Tokyo, Kashiwa, Chiba 277-8581, Japan
[8]Department of Physics and Astronomy, University of Missouri, Columbia, Missouri 65211-7010, USA
[#]E-mails: jgcheng@iphy.ac.cn, singhdj@missouri.edu

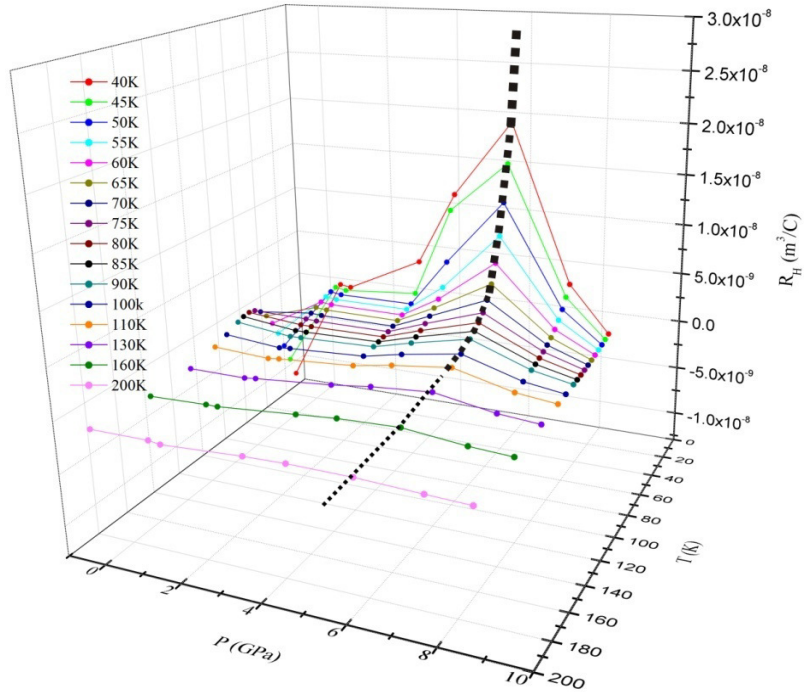

Fig. S1. A 3D plot of the Hall coefficient $R_H(T, P)$ highlighting the diverging behavior of $R_H$ upon cooling at 6.3 GPa if the superconducting transition is absent

16